\documentstyle[aps,multicol,tabularx,rotate,epsf]{revtex}
\begin{document}

\title{Electrostatic Contribution to Twist Rigidity of DNA}

\author{Farshid Mohammad-Rafiee and Ramin Golestanian}

\address{Institute for Advanced Studies in Basic Sciences, Zanjan 45195-159, Iran}

\date{\today}

\maketitle
\begin{abstract}
The electrostatic contribution to twist rigidity of DNA is
studied, and it is shown that the Coulomb self-energy of the
double-helical sugar-phosphate backbone contributes considerably
to twist rigidity of DNA---the electrostatic twist rigidity of DNA
is found as $C_{\rm elec}\approx 5$ nm, which makes up about $7
\%$ of its total twist rigidity ($C_{\rm DNA}\approx 75$ nm). The
electrostatic twist rigidity is found, however, to only weakly
depend on the salt concentration, because of a competition between
two different screening mechanisms: (1) Debye screening by the
salt ions in the bulk, and (2) structural screening by the
periodic charge distribution along the backbone of the helical
polyelectrolyte. It is found that depending on the parameters, the
electrostatic contribution could stabilize or destabilize the
structure of a helical polyelectrolyte.

\medskip
\noindent Pacs numbers: 87.15.-v, 36.20.-r, 61.41.+e
\end{abstract}
\pacs{87.15.-v,36.20.-r,61.41.+e}

\begin{multicols}{2}

\section{INTRODUCTION}  \label{sec:intro}

Genetic information in living cells is carried in the
double-helical linear sequence of nucleotides in DNA. The DNA
double-helix can be found in several forms that differ from each
other in the geometrical characteristics such as diameter and
handedness. Under normal physiological conditions, DNA adopts the
B-form, in which it consists of two helically twisted
sugar-phosphate backbones with a diameter $2.4$ nm, which are
stuffed with base pairs and are located asymmetrically with
respect to each other as characterized by the presence of major
and minor grooves. The helix is right-handed with $10$ base pairs
per turn, and the pitch of the helix is $3.4$ nm \cite{alb}.

It is well known that above pH 1 each phosphate group in DNA has a
negative charge \cite{bloomfield}, which renders the polymer stiff
due to the electrostatic repulsion between these groups. The
presence of neutralizing counterions and salt in the solvent
screens the electrostatic repulsion, thereby leading to an
effective way of controlling the stiffness of polyelectrolytes via
the ionic strength of the solution. To account for the
electrostatic stiffening, Odijk \cite{Odijk}, and Skolnick and
Fixman \cite{fixman} have adopted an effective wormlike chain
description for the bending elasticity of stiff polyelectrolytes,
and have calculated the correction to the persistence length due
to the electrostatic interactions. The so-called electrostatic
persistence length is found to be proportional to the square of
the Debye screening length, implying that the stiffness of
polyelectrolytes such as DNA should be very sensitive to salt
concentration \cite{Barrat1}. While there are experiments that
measure the electrostatic contribution to the bending rigidity of
DNA in various salt concentrations \cite{experiment1,experiment2},
it is generally believed that changing the ionic strength has no
significant effect on the rigidity of DNA under most
physiologically relevant conditions \cite{Hagerman1}. Hence, in
this so-called salt saturation limit, the bending rigidity of DNA
is entirely due to the mechanical stiffness of the double-helical
backbone.

\begin{figure}
\centerline{\epsfxsize=2.0cm\epsfbox{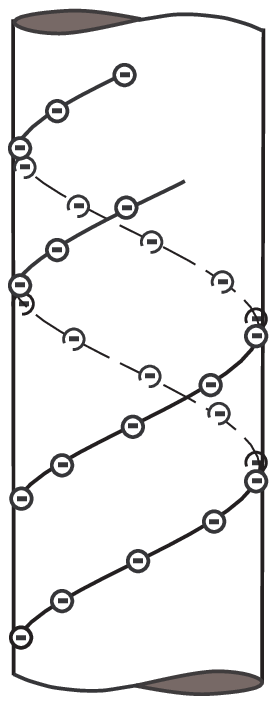}}
\vskip0.9truecm
\caption{The schematic picture of double-helical
B-DNA with the negative charges lying on the sugar-phosphate
backbone in a periodic manner.} \label{fig:schematics}
\end{figure}

\begin{figure}
\centerline{\epsfxsize 11cm {\epsffile{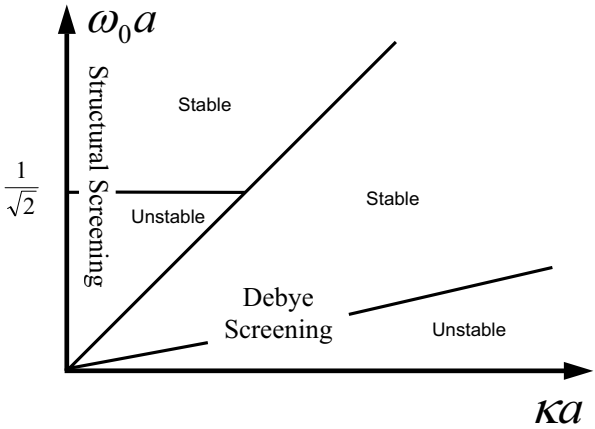}}} \caption{The
diagram delineating the different regimes in the parameter space
of a helical polyelectrolyte, where $\omega_0=2 \pi/P$, and $a$ is
the radius of the helix. The line separating the two screening
regimes has slope one, whereas the slope of the boundary denoting
the onset of instability in the Debye screening regime is set by
the inverse of a cutoff number $n_{\rm c}$ (see below).}
\label{fig:diagram}
\end{figure}

Similar studies have shown that the twist rigidity of DNA is also
relatively insensitive to the ionic strength of the solution
\cite{Hagerman2}. This experimental finding is usually explained
by saying that (unlike bending) twisting a polyelectrolyte does
not change the distance between the different charges on its
backbone appreciably, and thus it is not affected by electrostatic
interactions \cite{Kamien}. Here, we set out to revisit this line
of argument and attempt to account for the above experimental
observation from a different point of view. We consider the
electrostatic self-interaction of the double-helical
sugar-phosphate backbone (see Fig.~\ref{fig:schematics}) and show
that the periodic arrangement of the charge distribution
effectively screens the electrostatic interaction with the
screening length given by the pitch of the DNA. In other words,
corresponding to such a periodic charge distribution, there are
two competing screening lengths: (1) the Debye screening length of
the bulk solution $\kappa^{-1}$ that is controlled by the ionic
strength, and (2) the period of the charge distribution $P$, and
it is the smaller of these two lengths that controls the range of
Coulomb interaction. We find that electrostatic interactions make
an appreciable contribution to the twist rigidity of DNA, although
it depends only weakly on the Debye screening length as long as
this length is larger than the DNA pitch. We study the effect of
various geometrical parameters such as the diameter of the
double-helix, the distance between the two helices, and the pitch,
as well as the Debye screening length, on the electrostatic
contribution to twist rigidity and show that it can be both
positive and negative depending on the values of these parameters.
The results are summarized in Fig. \ref{fig:diagram}, where a
diagram is sketched in the parameter space delineating all the
different regimes.

The rest of paper is organized as follows: Section \ref{sec:model}
describes the model that is used to study the electrostatic
contribution to twist rigidity of DNA, followed by the
presentation of the results in Sec. \ref{sec:results}. Finally,
Sec. \ref{sec:discussion} concludes the paper, while some details
of the calculations appear in three Appendices.

\section{THE MODEL}     \label{sec:model}

To study the effect of electrostatic interactions on the twist
rigidity of DNA, we consider a simple model in which the
sugar-phosphate charged backbone of each DNA strand is assumed to
wrap around a cylinder of radius $a$ in a helical manner, as shown
in Fig. \ref{fig:schematics}. The double-helix can then be viewed
as a cylinder with a surface charge density $\sigma(z,\theta)$
corresponding to the negative charges, whose electrostatic
self-energy can be calculated as
\begin{eqnarray}
E_{\rm elec}&=&\frac{a^2}{2} \int d z d z^\prime \int_0^{2 \pi } d
\theta d \theta^\prime \; \sigma(z,\theta) \;
\sigma(z^\prime,\theta^\prime) \nonumber\\
&&\times \;\; V_{\rm
DH}\left(|\vec{r}(z,\theta)-\vec{r}(z^\prime,\theta^\prime)|\right),
\label{E-electro}
\end{eqnarray}
where $\vec{r}(z,\theta)$ parameterizes the position on the
surface of the cylinder with $z$ being the coordinate along the
axis and $\theta$ being the polar angle. The effective pair
potential between two charges in the solution is given by the
Debye-H\"{u}ckel interaction \cite{Debye}
\begin{equation}
V_{\rm DH}(r)=k_{\rm B} T \frac{\ell_{\rm B}}{r} {\rm e }^{-\kappa
r}, \label{DH}
\end{equation}
where $\ell_{\rm B}=e^2/(\epsilon k_{\rm B} T)$ is the Bjerrum
length, and $\kappa^{-1}$ is the Debye screening length, defined
via \cite{Barrat1}
\begin{equation}
\kappa^2=4 \pi \ell_{\rm B} \sum_i Z_i^2 c_i,
\label{concentartion}
\end{equation}
where $Z_i$ and $c_i$ are the valence and the concentration of the
salt species $i$, respectively, and summation holds over the ionic
species in the solution.

Due to the helical structure of DNA, $\sigma(z,\theta)$ is a
doubly periodic function, namely,
$\sigma(z,\theta)=\sigma(z+P,\theta)=\sigma(z,\theta+2\pi)$, where
$P$ is the helix pitch. Therefore, it is convenient to write
$\sigma(z,\theta)$ in the Fourier space as
\begin{eqnarray}
\sigma(z,\theta)=\sum_{m,n} \sigma_{mn} e^{i\frac{2\pi m}{P}z +
in\theta} \label{sigma1}
\end{eqnarray}
where, $m$ and $n$ are integer numbers.

\begin{figure}
\centerline{\epsfxsize=7.0cm\epsfbox{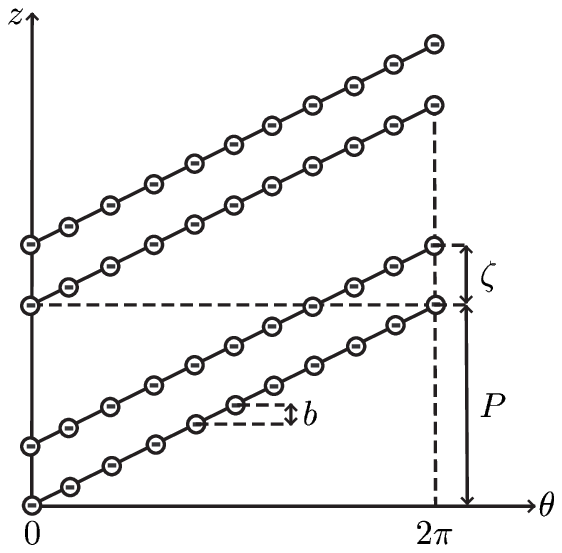}}
\vskip0.9truecm
\caption{The schematic picture of the surface charge distribution
of B-DNA. The geometrical parameters of DNA have been shown in the
picture: $b$ is the vertical distance between two successive
charges on each strand, $\zeta$ is the distance between the two
strands along the $z$-axis (given by the width of the minor groove
in B-DNA), and $P$ is the pitch of the helix.} \label{fig:charge}
\end{figure}

Making use of the periodicity of the charge distribution, one can
simplify the form of the electrostatic self-energy of Eq.
(\ref{E-electro}) using the Fourier representation of the screened
Coulomb interaction. After some manipulation, whose details can be
found in Appendix \ref{app:coulomb}, one finds
\begin{eqnarray}
\beta E_{\rm elec}&=& 4 \pi^2 \ell_{\rm B} L a^2 \sum_{m,n}
\left|\sigma_{m,n}\right|^2 I_n\left(\sqrt{(\kappa
a)^2+\left(n a \omega_0\right)^2}\right)\nonumber \\
&\times&  K_n\left(\sqrt{(\kappa a)^2+\left(n a
\omega_0\right)^2}\right), \label{Efinal1}
\end{eqnarray}
where $\beta=1/(k_{\rm B} T)$, $\omega_0=2\pi/P$ is the
spontaneous twist of the helix, and $L$ is the overall length of
the macromolecule.

We now focus on the specific case of DNA, whose charge density
$\sigma(z,\theta)$ can be written as (see Fig. \ref{fig:charge})
\begin{eqnarray}
\sigma(z,\theta)=-\frac{P}{2\pi a
b}\left[\delta(z-\frac{P\theta}{2\pi})+\delta(z-\zeta-\frac{P\theta}{2\pi})\right],
\label{sigma2}
\end{eqnarray}
where $b$ is the vertical distance between two charges in a
strand, and $\zeta$ is the distance between the two strands along
the $z$ direction, as shown in Fig. \ref{fig:charge}. Note that
\begin{eqnarray}
\int_0^P dz \int_0^{2\pi} a d\theta \;
\sigma(z,\theta)=-\frac{2P}{b}, \label{normalization}
\end{eqnarray}
yields the number of charges in each repeat unit of DNA. The
Fourier transform of the charge density $\sigma_{m, n}$ can now be
calculated from Eqs. (\ref{sigma1}) and (\ref{sigma2}) as
\begin{equation}
\sigma_{m,n}=-\frac{1}{2\pi ab}\delta_{m,-n} \left[1+e^{i n
\omega_0 \zeta}\right],  \label{sigmamn}
\end{equation}
using which the electrostatic self-energy of the double-helical
DNA can be calculated (from Eq. (\ref{Efinal1})) as
\begin{eqnarray}
\beta E_{\rm elec}&=&\frac{4\ell_{\rm B} L}{b^2}\sum_{n=0}^{\infty
\; \prime} \left(1+\cos n \omega_0 \zeta\right)
I_n\left(\sqrt{(\kappa a)^2+\left(n a
\omega_0\right)^2}\right) \nonumber \\
&&\times \; K_n\left(\sqrt{(\kappa a)^2+\left(n a
\omega_0\right)^2}\right), \label{Efinal}
\end{eqnarray}
where the prime indicates that the $n=0$ term should be counted
with a prefactor of $1/2$.

To calculate the contribution to twist rigidity from the above
Coulomb interaction, we impose an additional uniform twist of
$\Omega$ in the double-helix and calculate the change in the
self-energy, i.e. $\beta E_{\rm elec}(\omega_0+\Omega)-\beta
E_{\rm elec}(\omega_0)$. Expanding the energy change in powers of
$\Omega$, we can then read off the electrostatic twist rigidity as
\begin{equation}
C_{\rm elec}=\frac{1}{L}\frac{\partial^2 \beta E_{\rm
elec}}{\partial \Omega^2}, \label{Celec-def}
\end{equation}
subject to the constraint that the relative positioning of the two
helical strands should not alter upon deformation, and thus we
should assume that the parameter $\zeta$ changes accordingly to
$\zeta^\prime$ such that $(\omega_0+\Omega) \zeta^\prime=\omega_0
\zeta$.

\section{THE RESULTS}   \label{sec:results}

Under normal physiological conditions, $\kappa \approx 1 \; {\rm
nm}^{-1}$ and the spontaneous twist of B-DNA is $\omega_0=1.85 \;
{\rm nm}^{-1}$. Since the closed form calculation of $C_{\rm
elec}$ from Eqs. (\ref{Efinal}) and (\ref{Celec-def}) is
cumbersome, we choose to expand the modified Bessel functions
$I_n$ and $K_n$ to forth order in $(\kappa a)/(n a \omega_0)$.
This approximation appears to yield sufficient accuracy for the
experimentally relevant range of parameters.

It is convenient to use the asymptotic forms of $I_n(n x)$ and
$K_n(n x)$ for sufficiently large $n$. We find that $I_n(n x)K_n(n
x)=\frac{1}{2n}\frac{1}{\sqrt{1+x^2}}+O(\frac{1}{n^{2+\delta}})$
with $\delta \geq 0$, and observe that to a good approximation one
can just use the relevant asymptotic forms of $I_n$ and $K_n$ for
$n \geq 2$, in calculating the electrostatic twist rigidity (see
Appendix \ref{app:asympt} for details). We find
\begin{eqnarray}
C_{\rm elec}&=&\frac{2\ell_{\rm B} a^2}{b^2}(1+\cos \omega_0
\zeta)\nonumber\\
&&\times\left[f_0(a\omega_0)-f_2(a\omega_0) (\kappa a)^2
+f_4(a\omega_0) (\kappa a)^4\right] \nonumber\\
&+&\frac{2\ell_Ba^2}{b^2} \frac{(2
a^2\omega_0^2-1)}{(a^2\omega_0^2+1)^{5/2}}\sum_{n=2}^{\infty}{1
\over n }(1+\cos n \omega_0 \zeta), \label{C1}
\end{eqnarray}
where $f_0(x)$, $f_2(x)$, and $f_4(x)$ are functions defined in
Appendix \ref{app:explicit}. In the summation term in Eq.
(\ref{C1}) above, where we have used the asymptotic forms of the
Bessel functions, no dependence on $\kappa$ remains and only the
structural parameters of DNA such as $a$ and $\omega_0$ enter. The
summation diverges as $1/n$, and needs to be regularized with a
cutoff for $n$, which can be estimated as $n_{\rm c}=2\pi a /t$,
where $t$ is set by the thickness of each strand. Then, Eq.
(\ref{C1}) can be written as
\begin{eqnarray}
C_{\rm elec}&=&\frac{2\ell_{\rm B} a^2}{b^2}(1+\cos\omega_0\zeta)\nonumber\\
&&\times\left[f_0(a\omega_0) - f_2(a\omega_0) (\kappa a)^2+
f_4(a\omega_0) (\kappa a)^4\right]
\nonumber \\
&+&
\frac{2\ell_{\rm B}a^2}{b^2}\frac{(2a^2\omega_0^2-1)}{(a^2\omega_0^2+1)^{5/2}}\nonumber\\
&&\times \left[\gamma+\ln\frac{n_{\rm c}}{2\sin(\omega_0 \zeta
/2)}-(1+\cos\omega_0\zeta)\right], \label{C}
\end{eqnarray}
where $\gamma=0.577216$ is the Euler's constant. Note that the
above result, as we have already mentioned, is only valid for
$\kappa < \omega_0$.

Let us first evaluate the overall magnitude of the electrostatic
twist rigidity, as given by Eq. (\ref{C}). For B-DNA, we have
$a=1.2 \; {\rm nm}$, $\omega_0=1.85 \; {\rm nm}^{-1}$, $b=3.4
\;{\rm \AA}$ and $\zeta=1.13 \; {\rm nm}$ \cite{bloomfield}, and
the Bjerrum length is given as $\ell_{\rm B}=7.1 \;{\rm \AA}$.
Using these parameters, we find $C_0=174 \;{\rm \AA}$, which is
relatively large. To estimate $n_{\rm c}$ for B-DNA, we use $t
\approx 5 \;{\rm \AA}$, which gives $n_{\rm c} \approx 15$. Using
these estimates and $\kappa \approx 1 \; {\rm nm}^{-1}$, Eq.
(\ref{C}) yields $C_{\rm elec}=46 \;{\rm \AA}$ at physiological
salt concentration.

\begin{figure}
\centerline{\epsfxsize=8cm\epsfbox{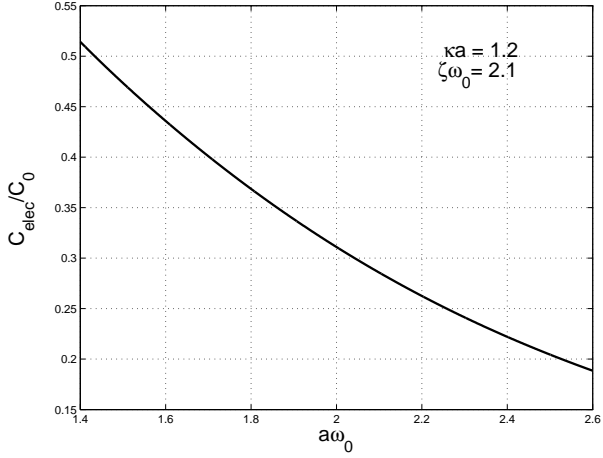}} \vskip0.5truecm
\caption{$C_{\rm elec}/C_0$ as a function of $a \omega_0$. This
plot corresponds to $\kappa a=1.2$ and $\zeta\omega_0=2.1$.}
\label{fig:aomega0-1}
\end{figure}

To study the effect of various parameters, namely, the spontaneous
twist, the diameter, and the asymmetry of the double-helix, as
well as the salt concentration, we choose to work with three
dimensionless parameters of $a \omega_0$, $\kappa a$, and $\zeta
\omega_0$. In Fig. \ref{fig:aomega0-1}, the behaviour of $C_{\rm
elec}$ is shown as a function of $a \omega_0$ for $\kappa a=1.2$
and $\zeta \omega_0=2.1$. The domain for $a \omega_0$ is chosen
such that the condition $\kappa < \omega_0$ is satisfied and Eq.
(\ref{C}) is valid. The plot shows that for sufficiently high salt
concentration, the electrostatic torsional stiffness decreases as
the spontaneous twist of the double-helix increases.

\begin{figure}
\centerline{\epsfxsize=8cm\epsfbox{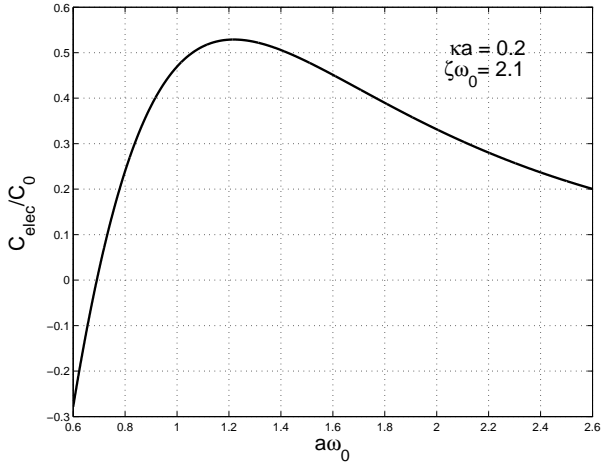}} \vskip0.5truecm
\caption{$C_{\rm elec}/C_0$ as a function of $a \omega_0$. This
plot corresponds to $\kappa a=0.2$ and $\zeta\omega_0=2.1$.}
\label{fig:aomega0-2}
\end{figure}

For sufficiently low salt concentration, however, it appears that
the behavior is not always monotonic as shown in Fig.
\ref{fig:aomega0-2}, where $C_{\rm elec}$ is plotted as a function
of $a \omega_0$ for $\kappa a=0.2$ and $\zeta \omega_0=2.1$.
Interestingly, one can see that $C_{\rm elec}$ can even become
negative, due to the fact in Eq. (\ref{C}) the first term becomes
relatively weak for low salt concentrations and the second term
that is dominant changes sign for $a \omega_0<1/\sqrt{2}$.

\begin{figure}
\centerline{\epsfxsize=8cm\epsfbox{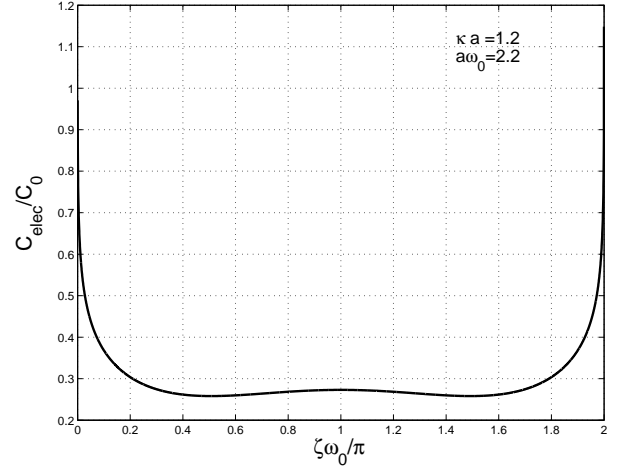}} \vskip0.5truecm
\caption{$C_{\rm elec}/C_0$ as a function of $\zeta \omega_0
/\pi$. This plot corresponds to $\kappa a=1.2$ and $a
\omega_0=2.2$.} \label{fig:zeta}
\end{figure}

In Fig. \ref{fig:zeta}, the dependence of $C_{\rm elec}$ is shown
on the asymmetry parameter $\zeta \omega_0$ for $\kappa a=1.2$ and
$a \omega_0=2.2$. One observes that for the relatively large
window of $0.4\pi \leq \zeta \omega_0 \leq 1.6\pi$, $C_{\rm elec}$
is almost constant, and is thus not sensitive to the relative
positioning of the two strands. For $\zeta\omega_0=0$ and $2\pi$,
a divergence sets in due to the fact that the charges on the two
strands develop contacts with each other.

\begin{figure}
\centerline{\epsfxsize=8cm\epsfbox{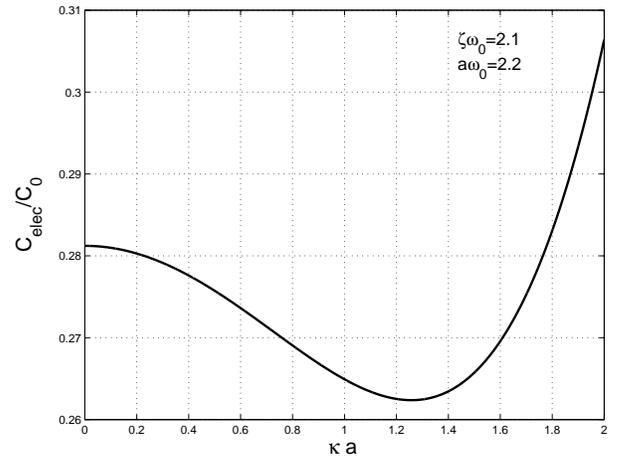}} \vskip0.5truecm
\caption{$C_{\rm elec}/C_0$ as a function of $\kappa a$. This plot
corresponds to $\zeta \omega_0=2.1$ and $a \omega_0=2.2$.}
\label{fig:kappa}
\end{figure}

Finally, the behavior of $C_{\rm elec}$ is shown in Fig.
\ref{fig:kappa} as a function of $\kappa a$ for
$\zeta\omega_0=2.1$ and $a\omega_0=2.2$. The dependence on the
Debye screening parameter in Eq. (\ref{C}) comes only from the
first term, where the negative sign of the coefficient of $(\kappa
a)^2$ causes a dip in the plot for $C_{\rm elec}$ around $\kappa a
\approx 1.3$, which by chance corresponds to the normal
physiological salt concentration. However, as can be seen from
Fig. \ref{fig:kappa}, the dependence of the electrostatic twist
rigidity on $\kappa a$ is extremely weak as long as $\kappa <
\omega_0$, which is a manifestation of the fact that screening is
controlled by the periodic charge distribution and the effective
screening length is set by the pitch $P$ that is shorter than
$\kappa^{-1}$ in this regime.

It is worth saying a few words about the other limit where $\kappa
> \omega_0$, corresponding to high salt concentration. In Eq.
(\ref{Efinal}), one can manifestly see that in the $n$th term in
the series there is a competition between $\kappa$ and $n
\omega_0$ to control the screening. If the salt concentration is
so high that we have $\kappa > n_{\rm c} \omega_0$, the periodic
structure plays no role and screening is entirely controlled by
the Debye screening in the bulk. For relatively strong Debye
screening when $\kappa a > 1$, we can use the simple asymptotic
forms of the Bessel functions and find an asymptotic expression
for the electrostatic twist rigidity as
\begin{eqnarray}
C_{\rm elec}&=&\frac{2\ell_{\rm B}a^2}{b^2}\sum_{n=1}^{n_{\rm c}}
(1+\cos n \omega_0 \zeta) \frac{n^2 [2(n a \omega_0)^2-(\kappa
a)^2]} {[(\kappa a)^2+(n a \omega_0)^2]^{5/2}}.\nonumber \\
\label{Csalt}
\end{eqnarray}
This expression can be used for the region $\kappa
> \omega_0$, where it predicts $C_{\rm elec}>0$
for $\kappa < s \omega_0$, and $C_{\rm elec}<0$ for $\kappa > s
\omega_0$, for a value of $s \approx n_{\rm c}$.

\section{Discussion}         \label{sec:discussion}

For a polyelectrolyte with a periodic spatial charge distribution,
such as the double-helical structure of DNA, there are two
competing mechanisms for screening the electrostatic
self-interaction and its contribution to twist rigidity, namely,
the {\it Debye screening} due to the free ions in the solution,
and the {\it structural screening} caused by the periodic
structure of the charge distribution. While the screening length
for the former case is set by the Debye length $\kappa^{-1}$, it
is set by the period $P$ of the charge distribution in the latter,
and the dominant mechanism corresponds to the one with the shorter
screening length.

It appears that the contribution of electrostatic interactions to
twist rigidity can be both negative and positive, depending on the
parameters. The negative values for the electrostatic torsional
stiffness could lead to instability in the structure of the
helical polyelectrolyte, depending on whether the mechanical
structure of the macromolecule can counter-balance the effect of
the electrostatic instability. We have used this criterion in Fig.
\ref{fig:diagram} to summarize these different regimes in the
parameter space. Considering that a helical polyelectrolyte seems
to be the general structure of many stiff biopolymers (such as DNA
and actin) it will be interesting to know which helical
configurations can in principle lead to stable structures, and
which ones cannot. This could especially be important in the case
of biopolymers that self-assemble through polymerization
processes, such as actin, where such energetic considerations
could hamper or favor the polymerization process.

The twist rigidity of B-DNA is believed to be roughly 75 nm
\cite{Vologodskii}, which should be thought of as the sum of the
mechanical and the electrostatic contributions: $C=C_{\rm
mech}+C_{\rm elec}$. Our estimate of $C_{\rm elec} \approx 5$ nm
reveals that about $7 \%$ of the twist rigidity of DNA is due to
electrostatic interactions. This result is more or less
independent of the salt concentration, as the dependence of
$C_{\rm elec}$ to salt concentration is very weak. For example,
the difference between $C_{\rm elec}$ at zero and very high salt
concentration is about $3.5 \;{\rm \AA}$. Therefore ${\Delta
C_{\rm elec}/C}=0.5 \%$, which is very small. While this naturally
explains why in the experiments no sensitivity on the salt
concentration has been observed, it certainly does not mean that
the electrostatic contribution to $C$ is negligible.

In the above analysis we have assumed that imposing a finite angle
of twist between the two ends of a helical polyelectrolyte leads
to a uniform twist. This is analogous to the assumption of a
uniform bending made by Odijk when calculating the electrostatic
bending rigidity, and presumably holds true when the effective
elasticity due to electrostatics is local. In Ref. \cite{Zandi},
this assumption has been scrutinized and it has been shown that
this assumption is valid provided one of these conditions hold:
(1) the polyelectrolyte segment is long, (2) the Debye screening
is strong, (3) the charging is weak, or (4) the mechanical
stiffness of the polyelectrolyte is larger than the electrostatic
contribution. We expect the same argument to hold true for the
twist rigidity as well. Since for the case of DNA we have shown
that the mechanical twist rigidity is much larger than the
electrostatic contribution, we can safely assume that the twist is
uniform.

In conclusion, we have studied the electrostatic contribution to
twist rigidity of DNA, taking into account its dependence on salt
concentration in the solvent. We have shown that there is a
non-negligible electrostatic contribution to twist rigidity, which
varies very slowly by changing the salt concentration in the
solution. By changing the geometrical parameters of the helix and
the Debye screening length, the electrostatic twist rigidity can
change sign and become negative, implying that a helical structure
could be a both stable as well as unstable configuration for a
helical polyelectrolyte. We finally note that the present analysis
can be also applied to other biopolymers such as F-actin.

\acknowledgments

We are grateful to T.B. Liverpool for interesting discussions and
comments.

\appendix

\section{Coulomb Energy in Fourier Space}       \label{app:coulomb}

Due to the periodicity of the charge distribution, it is
convenient to calculate the electrostatic self energy in Eq.
(\ref{E-electro}) in Fourier space. We start from the Fourier
representation of the screened Debye-H\"{u}ckel interaction in Eq.
(\ref{DH}):
\begin{eqnarray}
\beta E_{\rm elec}=\frac{a^2}{2}\int \frac{d^3k}{(2\pi)^3} \int &
d z & d z^\prime \int_{0}^{2 \pi} d \theta d \theta^\prime
\; \sigma(z,\theta)\sigma(z^\prime,\theta^\prime) \nonumber\\
&\times&  \frac{4\pi \ell_B}{k^2+\kappa^2}~
e^{i\vec{k}\cdot\left[\vec{r}(z,\theta)-\vec{r}(z^\prime,\theta^\prime)\right]}.
\label{E1}
\end{eqnarray}
Using the cylindrical coordinates, we can write $\vec{r}(z,\theta)
= z \hat{z} + a (\cos \theta \; \hat{x}+ \sin \theta \; \hat{y})$,
which in conjunction with $\vec{k}= k_z \hat{z} + k_\bot (\cos
\phi \;\hat{x}+\sin \phi \; \hat{y})$, yields
\begin{eqnarray}
\beta E_{\rm elec}&=&\frac{\ell_{\rm B} a^2}{2}\int dz dz^\prime
\int_{0}^{2\pi} d\theta d\theta^\prime \nonumber \\
&\times& \int_{-\infty}^{+\infty}\frac{dk_z}{2\pi}
\int_{0}^{\infty}
\frac{k_\bot dk_\bot}{(2\pi)^2} \int_{0}^{2\pi} d \phi \nonumber \\
&\times& \sum_{m,n} \sum_{m^\prime,n^\prime}\sigma_{m,n}
\sigma_{m^\prime,n^\prime} \frac{4\pi}{k_z^2+k_\bot^2+\kappa^2}
\nonumber \\
&\times& e^{i\left(\frac{2\pi}{P}m+k_z\right)z +in\theta+ik_\bot a
\cos(\phi-\theta)}
\nonumber \\
&\times& e^{i\left(\frac{2\pi}{P}m^\prime-k_z\right)z^\prime +
in^\prime \theta^\prime - i k_\bot a \cos(\phi-\theta^\prime)}.
\label{Ez}
\end{eqnarray}
After integration over $z$, $z^\prime$, and $k_z$, we find
\begin{eqnarray}
\beta E_{\rm elec}&=&\frac{\ell_{\rm B} a^2}{2} \int_{0}^{2\pi}
d\theta d\theta^\prime  d\phi \int_{0}^{\infty} \frac{k_\bot
dk_\bot}{(2\pi)^2} \nonumber \\
&&\times \sum_{m,n}\sum_{m^\prime,n^\prime}\sigma_{m,n}
\sigma_{m^\prime,n^\prime} \frac{4\pi L
\delta_{m,-m^\prime}}{(\frac{2\pi}{P}m)^2+k_\bot^2+\kappa^2}\nonumber \\
&&\times \; e^{i(n\theta+n^\prime\theta^\prime) + i a k_\bot
[\cos(\phi-\theta)-\cos(\phi-\theta^\prime)]}. \label{Ephi}
\end{eqnarray}
By defining $\theta=\theta_1+\phi$ and
$\theta^\prime=\theta_2+\phi$, the integration over the three
angles in Eq. (\ref{Ephi}) can be performed as
\begin{eqnarray}
\int d\theta d\theta^\prime d\phi e^{i(n\theta+n^\prime
\theta^\prime)+iak_\bot[\cos(\phi-\theta)-\cos(\phi-\theta^\prime)]}=\nonumber\\
(2\pi)^3 \delta_{n,-n^\prime} [J_n(k_\bot a)]^2, \label{angle}
\end{eqnarray}
to yield
\begin{eqnarray}
\beta E_{\rm elec}&=& 4 \pi^2 \ell_{\rm B} L a^2 \sum_{m,n}
\left|\sigma_{m,n}\right|^2 \nonumber \\
&&\times\int_0^\infty dk_\bot k_\bot \frac{J_n^2(k_\bot
a)}{(\frac{2\pi}{P}m)^2+k_\bot^2+\kappa^2}. \label{EJn}
\end{eqnarray}
Performing the final integration over $k_\bot$ using \cite{Table}
\begin{equation}
\int_0^\infty \frac{x}{x^2+h^2}\left[J_\nu(x)\right]^2 dx =
I_\nu(h)K_\nu(h),   \label{integral}
\end{equation}
we obtain the result quoted in Eq. (\ref{Efinal1}).

\section{Asymptotic forms of the Bessel functions}       \label{app:asympt}

In this Appendix, the asymptotic forms of $I_n(nx)$ and $K_n(nx)$,
for large $n$ are derived. We use the integral representation of
these functions:
\begin{eqnarray}
I_n(nx)&=&\frac{1}{\pi^{\frac{1}{2}}(n-\frac{1}{2})!}
\left(\frac{nx}{2}\right)^n \int_{-1}^{+1} d p \; e^{nxp}
\left(1-p^2\right)^{n-\frac{1}{2}}  \nonumber\\
\label{I} \nonumber \\
K_n(nx)&=&\frac{\pi^\frac{1}{2}}{\left(n-\frac{1}{2}\right)!}
\left(\frac{nx}{2}\right)^n \int_1^{\infty} d p \;
e^{-nxp}\left(p^2-1\right)^{n-\frac{1}{2}}. \nonumber\\
\label{K}
\end{eqnarray}
Let us first consider the case of $I_n(nx)$. We denote
\begin{eqnarray}
Q&\equiv&\int_{-1}^{+1} d p \; e^{nxp}
\left(1-p^2\right)^{n-\frac{1}{2}}=\int_{-1}^{+1} d p \; e^{g(p)},
\label{Q}
\end{eqnarray}
and expand $g(p)$ around $p_0$, the position of its maximum, to
second order of $(p-p_0)$, as
\begin{eqnarray}
g(p)&\simeq&g(p_0)+\frac{1}{2}f^{\prime\prime}(p_0)(p-p_0)^2\nonumber\\
&=&nxp_0+\left(n-\frac{1}{2}\right)\ln
\left(1-p_0^2\right)\nonumber\\
&&-\left(n-\frac{1}{2}\right)\frac{1+p_0^2}{\left(1-p_0^2\right)^2}
(p-p_0)^2, \label{gp}
\end{eqnarray}
where
\begin{equation}
p_0=\frac{-1+\sqrt{1+x^2}}{x}+\frac{1}{2nx}\left(1+\frac{1}{\sqrt{1+x^2}}\right).\label{p0}
\end{equation}
Using this form for $g(p)$, $Q$ can be found as
\begin{eqnarray}
Q \simeq \frac{1-p_0^2}{\sqrt{n\left(1+p_0^2\right)}}
~e^{nxp_0+\left(n-\frac{1}{2}\right)\ln \left(1-p_0^2\right)}.
\end{eqnarray}
Finally, using the Stirling's formula for $n!$, we find
\begin{eqnarray}
I_n(nx)&\simeq& \frac{1}{\sqrt{2}\pi}\frac{1-p_0^2}
{\sqrt{n\left(1+p_0^2\right)}}\nonumber\\
&\times& e^{\left(n-\frac{1}{2}\right)\ln
\left(1-p_0^2\right)+nxp_0+n\left(1+\ln
\frac{x}{2}\right)+\frac{1}{24n}}.\label{In}
\end{eqnarray}
Using a similar treatment, we find the large $n$ asymptotic
behavior for $K_n(nx)$, as
\begin{eqnarray}
K_n(nx)&\simeq& \frac{1}{\sqrt{2n}}\frac{p_0^{\prime ~ 2}-1}
{\sqrt{n\left(1+p_0^{\prime ~ 2}\right)}}\nonumber\\
&\times& e^{\left(n-\frac{1}{2}\right)\ln \left(p_0^{\prime ~
2}-1\right)-nxp^\prime_0+n\left(1+\ln
\frac{x}{2}\right)+\frac{1}{24n}},\label{Kn}
\end{eqnarray}
where
\begin{equation}
p_0^\prime=\frac{1+\sqrt{1+x^2}}{x}-\frac{1}{2nx}
\left(1+\frac{1}{\sqrt{1+x^2}}\right).\label{p0'}
\end{equation}
By using these relations for $I_n(nx)$ and $K_n(nx)$, we find
\begin{eqnarray}
I_n(nx)K_n(nx)=\frac{1}{2\sqrt{1+x^2}}\frac{1}{n}+O(\frac{1}{n^{2+\delta}}),
\label{InKn}
\end{eqnarray}
where $\delta \geq 0$.

\begin{figure}
\centerline{\epsfxsize=8cm\epsfbox{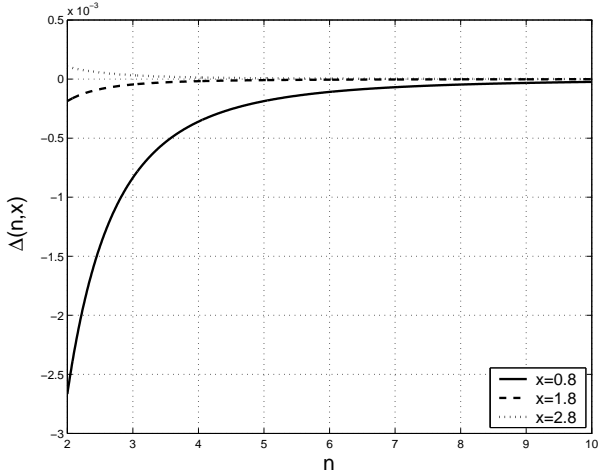}}
\caption{Difference between the asymptotic form of
$I_n(nx)K_n(nx)$ and its exact value as a function of $n$ for
several values of $x$. The solid line corresponds to $x=0.8$, the
dashed line corresponds to $x=1.8$, and the dotted line
corresponds to $x=2.8$. The difference is less than $0.25\%$.}
\label{fig:asymp1}
\end{figure}

We define $\Delta(n,x) \equiv
I_n(nx)K_n(nx)-\frac{1}{2n\sqrt{1+x^2}}$, and in Fig.
\ref{fig:asymp1} show the behaviour of $\Delta(n,x)$ as a function
of $n$ for several values of $x$. As can be seen, the difference
is less than $0.25 \%$  in the worst case, which implies that the
asymptotic form of $I_n(nx)K_n(nx)$ for $n \geq 2$ can be used as
a good approximation for the range of $x$ we are interested in.

\begin{figure}
\centerline{\epsfxsize=8cm\epsfbox{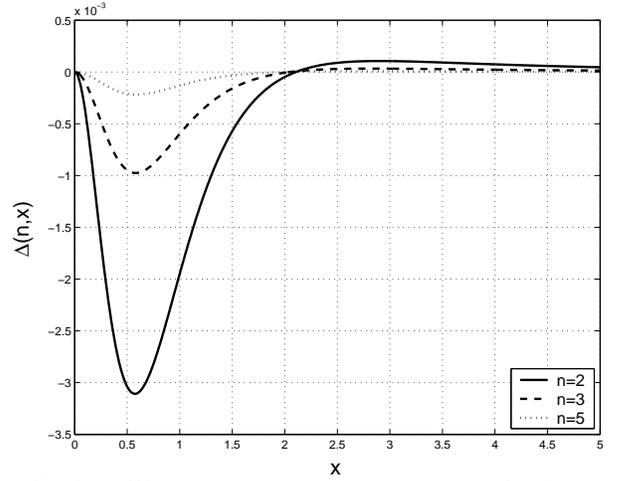}}
\caption{Difference between the asymptotic form of
$I_n(nx)K_n(nx)$ and its exact value as a function of $x$ for
different values of $n$. The solid line corresponds to $n=2$, the
dashed line corresponds to $n=3$, and the dotted line corresponds
to $n=5$. The difference is less than $0.3\%$.} \label{fig:asymp2}
\end{figure}

In Fig. \ref{fig:asymp2}, we show the behaviour of $\Delta(n,x)$
as a function of $x$ for different values of $n$. This plot shows
that for $n \geq 2$, the difference goes to zero as $x$ increases.

\section{The Explicit Forms of the Auxiliary functions}     \label{app:explicit}

In this Appendix, we give the explicit forms of the auxiliary
functions used in Eqs. (\ref{C1}) and (\ref{C}) above. The
function $f_0(x)$ reads
\begin{eqnarray}
f_0(x)&=&4 \left(1+\frac{1}{x^2}\right)I_1(x)K_1(x)
+4 I^\prime_1(x)K^\prime_1(x)\nonumber\\
&-&\frac{2}{x}\left[I_1(x)K^\prime_1(x)+I^\prime_1(x)K_1(x)
\right], \label{f0}
\end{eqnarray}
where the prime indicates differentiation. One can show that for
$x\ll 1$ it behaves as $f_0(x)=\ln x+O(x^2\ln x)$. The second
function $f_2(x)$ is written as
\begin{eqnarray}
f_2(x)&=& -3 \left(\frac{1}{x}+\frac{1}{x^3}\right)
\left[I_1(x)K^\prime_1(x)+I^\prime_1(x)K_1(x)\right]
\nonumber \\
&+&\frac{1}{x}\left[I^\prime_1(x)K^\prime_2(x)-I^\prime_2(x)K^\prime_1(x)\right]
\nonumber \\
&+&\frac{3}{x^2}\left[I_1(x)K_1(x)-2I^\prime_1(x)K^\prime_1(x)\right]
\nonumber \\
&+&\frac{3}{2x^2}\left[I^\prime_2(x)K_1(x)-I_1(x)K^\prime_2(x)\right]
\nonumber \\
&+&\frac{2}{x^3}\left[I_1(x)K_2(x)-I_2(x)K_1(x)\right],
\end{eqnarray}
and for $x\ll 1$ it behaves as $f_2(x)=1/(2x^2)-\ln x/2 +O(x^2\ln
x)$. Finally, for $f_4(x)$ we have
\begin{eqnarray}
f_4(x)&=&\frac{176+136x^2+19x^4}{16 x^6}I_1(x)K_1(x)
\nonumber \\
&-&\frac{7}{8x^3}\left[I_3(x)K^\prime_1(x)+I^\prime_1(x)K_3(x)\right]
\nonumber \\
&+&\frac{1}{x^4}\left[I_2(x)K^\prime_1(x)-I^\prime_1(x)K_2(x)\right]
\nonumber\\
&+& \frac{11+2x^2}{x^4}I^\prime_1(x)K^\prime_1(x)+
\frac{1}{16x^2}I_3(x)K_3(x) \nonumber \\
&+&\frac{14+3x^2}{8x^4}\left[I^\prime_2(x)K_1(x)-I_1(x)K^\prime_2(x)\right]
\nonumber \\
&-&
\frac{88+49x^2}{8x^5}\left[I^\prime_1(x)K_1(x)+I_1(x)K^\prime_1(x)\right]
\nonumber \\
&-& \frac{1}{x^5}\left[I_2(x)K_1(x)-I_1(x)K_2(x)\right],
\end{eqnarray}
which behaves as $f_4(x)=3/(4x^4)-1/(8x^2)+5/64\ln x+O(x^2\ln x)$,
for $x\ll 1$. At infinity, all of these functions go to zero
faster than $1/x^2$, with $f_4(x)$ vanishing faster than $f_2(x)$,
and $f_2(x)$ faster than $f_0(x)$, respectively.

\begin{figure}
\centerline{\epsfxsize=8cm\epsfbox{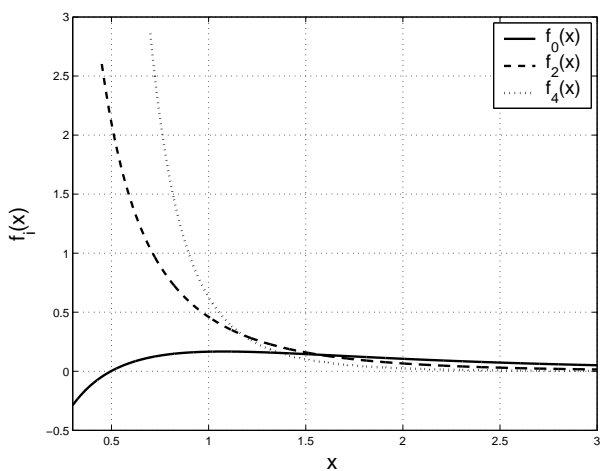}} \caption{The
auxiliary functions $f_0(x)$, $f_2(x)$, and $f_4(x)$.}
\label{fig:fi(x)}
\end{figure}

\end{multicols}
\end{document}